\documentclass[manuscript]{aastex}

\newcommand{\gapprox}{$\stackrel {>}{_{\sim}}$}   

\shorttitle{PV Cep IR variability}
\shortauthors{Lorenzetti et al.}

\begin{document}

\title{Mid- and Far-Infrared Variability of PV Cep}

\author{D.Lorenzetti\altaffilmark{1}, S.Antoniucci\altaffilmark{1}, T. Giannini\altaffilmark{1}, G.Li Causi\altaffilmark{1}, A. Di Paola\altaffilmark{1}}
\affil{INAF - Osservatorio Astronomico di Roma - Via Frascati 33\\ 00040 Monte Porzio Catone, Italy}

\author{A.A. Arkharov\altaffilmark{2}}
\affil{Central Astronomical Observatory of Pulkovo - Pulkovskoe shosse 65, 196140 St.Petersburg, Russia}

\and

\author{V.M. Larionov\altaffilmark{2,3}}
\affil{Astronomical Institute of St.Petersburg University, Russia}


\begin{abstract}
We present the collection of all the mid- and far-IR observations ($\lambda$=3-170$\mu$m) of the young eruptive variable PV Cep available so far in the
literature. These data allow us to confirm that flux variability is a prominent feature at mid-IR wavelength ($\lambda$=3-25$\mu$m).
Color-magnitude plots clearly indicate that the observed variability is not extinction-driven, but mainly
influenced by fluctuations of the mass accretion rate. We interpret such variability as due to a hot spot created onto the stellar surface by the column of accreting matter, which heats the inner parts of the disk and determines the observed increase of the near- mid-IR luminosity. A quantitative characterization is given for both the spot itself and the additional thermal component created by it. Far-IR data ($\lambda$=60-170$\mu$m) are consistent with the presence of a temperature stratification in a massive and quite un-evolved circumstellar disk.
\end{abstract}


\section{Introduction}

After the very early phases of the cloud collapse the formation of an accretion disk around the central protostar
is a natural consequence of both the initial cloud rotation and the angular momentum conservation. Once the disk is
formed the accretion can proceed by transferring angular momentum toward the external regions and matter toward the 
central object through the disk itself. This is not a continuous but an intermittent process
characterized by episodes during which matter from the inner disk regions falls onto the protostar following 
its magnetic field interconnection lines (e.g. Shu et al. 1994). The accreting matter
produces a shock that cools by emitting a hot continuum. Moreover, as a consequence of the accretion event, 
strong winds (in some cases also collimated jets) emerge from the rotating star/disk system. 

Disk accretion phenomena (during which the mass accretion rate increases up to orders of magnitude)
give origin to repetitive outbursts (Hartmann \& Kenyon 1985) that are typical of many (if not all) young stellar objects (YSO's). Recently, D'Angelo \& Spruit (2010) provided quantitative predictions for the episodic accretion on magnetized stars and indication that the cycle time of the bursts increases with a decreasing accretion rate.
The outbursts of largest intensity (\gapprox~4 mag) are classified into two major classes: ({\it i}) EXor events (Herbig 1989) lasting one year or less, with a recurrence time of months to years, characterized by emission line spectra;
({\it ii}) FUor events (Hartmann \& Kenyon 1985) of longer duration (\gapprox~tens of years)
with spectra dominated by absorption lines.\ 

Circumstellar disks survive few million years before dissipating and undergo different phases of evolution: from accretion, through transitional, to passive disks, that make them crucial structures to investigate accretion modalities, matter outflowing, disk excavation and planet formation. Detailed studies on the correlations between optical, near-, and mid-infrared (IR) variability of protostellar objects are very important since they can elucidate on the physical mechanism(s) responsible for the fluctuations themselves. Of particular interest is the mid-IR domain (3-25 $\mu$m) where many of the eruptive variables emit most of their energy (Antoniucci et al. 2013). The spectral behaviour at these wavelengths is strictly related to disk and envelope regions located at radial distances from the central star  where disk fragmentation and planet formation may occur. The difficult access to facilities operating in this spectral range has hampered this kind of studies, apart few important exceptions (eg. Kun et al. 2011, hereinafter K11; K\'{o}sp\'{a}l et al. 2012 and references therein; Antoniucci et al. 2013).\

One of the few eruptive YSOs observed in the mid- far-IR in several occasions is PV Cephei ($\alpha_{2000}$ = 20$^{h}$45$^{m}$53.96$^{s}$, $\delta_{2000}$ =  +67$^{\circ}$57$^{\prime}$38.9$^{\prime \prime}$), which
is a pre-main sequence star in the northeastern edge of the L1158 and L1155 groups of dark clouds, at a distance of 325 pc
(Straizys et al. 1992). Its strong and irregular variability (up to about 4-5 mag) was recently studied in the optical and near-IR bands [K11; Lorenzetti et al. 2011 (Paper I), 2013].
The aims of the present work are twofold: ({\it i}) to provide a mid- far-IR database of PV Cep, collecting all the observations obtained so far;
({\it ii}) to understand whether photometric variations at longer wavelengths present some analogies with those occurring in the optical/near-IR range
in which case they could originate from the same mechanism(s).
Sect. 2, summarizes the results of the recent optical/near-IR monitoring. In Sect.3 the mid-and far-IR data are presented and discussed in Sect.4. 
Sect.5 summarizes our conclusions.

\section{PV Cep NIR-IR variability}

As mentioned above, K11 and Paper I independently report on the optical/near-IR variability of PV Cep since 2004. The optical/near-IR variability of PV Cep is firmly ascertained and its origin is often attributed to a variable accretion rate, albeit some aspects remain still open. K11 suggest that the large and spectacular fading (reaching 4 mag in I$_C$ band) started in 2005, can not be due only to the accretion rate decrease (responsible, at most of about 1 mag), but also to a substantial extinction variation. 
Their interpretation, although correct in a general sense, should be applied with some caveats to this specific case. Indeed, ({\it i}) accretion rates
are empirically derived from the Ca~II infrared triplet emission lines, whose diagnostic capability suffer from some uncertainties (e.g. Antoniucci et al. 2011); ({\it ii}) the magnitude variation as a function of the [R-I] color (Figure 2 of K11) is surprisingly almost orthogonal to the extinction vector.
In Paper I the presented behaviour of PV Cep refers to a period of less intense activity in the optical and near-IR range. Noticeably, both the near-IR and optical col-col plots (Figures 2 and 3 of Paper I) indicate that, beside a constant extinction A$_V$ of about 5-7 mag, phenomena other than extinction have a significant role in producing all the observed fluctuations. Admittedly, the near-IR color variations could be misinterpreted as due to extinction variations with A$_V$ values between 7 and 20 mag, but the optical colors demonstrate that this visual extinction range is definitely inconsistent. 

\section{PV Cep mid-and far-IR photometry} 

To ascertain if PV Cep presents significant fluctuations in the mid- far-IR range, as it certainly does in the 
optical/near-IR, we have collected in Table~\ref{variab:tab} all the observations available in the literature taken in the L 
($\sim$3$\mu$m), M ($\sim$5$\mu$m), N ($\sim$10$\mu$m), and Q ($\sim$20$\mu$m) bands and at 60, 100, and 170 $\mu$m. 
A similar kind of analysis, essentially based on {\it Spitzer} data  taken in different epochs, was already presented by K11.
The selected bands are those where photometry has been performed in at least three occasions.
To construct the table we have adopted all the photometry obtained with ground-based facilities, IRAS (12, 25, 60 and 100 $\mu$m), {\it Spitzer} 
(IRAC 3.4 and 4.5 $\mu$m; MIPS 24, 70 $\mu$m), and {\it WISE} (3.6, 4.6, 12 and 22 $\mu$m). No counterpart is given in MSX Catalogue.
Data taken in different, although adjacent, bands (eg. IRAC 5.8 and 8.0 $\mu$m, AKARI 9 $\mu$m) have not been considered because 
of the large difference between the effective wavelengths and band-passes. 
We signal that in 1987 PV Cep resulted too weak to be detected by  {\it KAO} (Kuiper Astronomical Observatory) at 50 and 100 $\mu$m (Natta et al. 1993).\

The epoch of each observation and the effective wavelengths of the used band-passes are also provided in Table~\ref{variab:tab}. All values are given as flux density (in Jansky); when necessary, the magnitude values found in the literature have been converted by adopting the zero-magnitude fluxes given in the web pages of the Gemini Observatory\footnote{http://www.gemini.edu/?q=node/11119}, whenever the conversion values are not given in the original paper. This procedure, together with the comparison between band-passes which are not perfectly coincident, is affected by instrumental and calibration effects and may cause a scatter that we conservatively estimate to be up to 50\%. Therefore, 
fluctuations up to that amplitude will be considered negligible in the present work, unless the measurements have been performed with the same instrumentation: in these latter cases the fluctuations are assumed to be significant if their amplitude is larger than the 3$\sigma$ uncertainty. In the far-IR, PV Cep was observed in different occasions and the resulting photometry is given in the last part of Table~\ref{variab:tab}.\

As for the WISE measurements, the average L and M bands magnitudes from the All Sky Source Catalog (2.43 Jy and 8.15 Jy, respectively, not reported in Table~1) show a flux variation of opposite sign in the two bands with respect to previous photometries. To verify the reliability of this result, we scrutinized the single WISE measurements and noted that these are characterized by a very high dispersion (around 3 Jy). This dispersion becomes even higher in later Post Cryo WISE observations (NEO WISE Catalog). Moreover, all the WISE data are significantly affected by saturation effects,
an inconvenient overcome by {\it Spitzer} data, obtained in High Dynamic Range (HDR) mode (i.e. co-adding very short exposures). Because of these reasons, we opted to discard the WISE L and M bands measurements in this work.

Far-IR fluxes have to be cautiously treated in any merging with other bands, because they are typically obtained with a beam size much larger than that of other observations. In particular, the IRAS measurements reported in the Table are not used in the following analysis because the large beam size affects especially the longest wavelength.
Photometry in the 25-100 $\mu$m range indicates a marginal degree of variability (mainly at the shorter wavelengths). This could be likely due to beam differences rather than to intrinsic fluctuations.

\section{Analysis and discussion}

\subsection{Mid-IR behaviour}

\begin{figure}
\includegraphics[width=14cm]{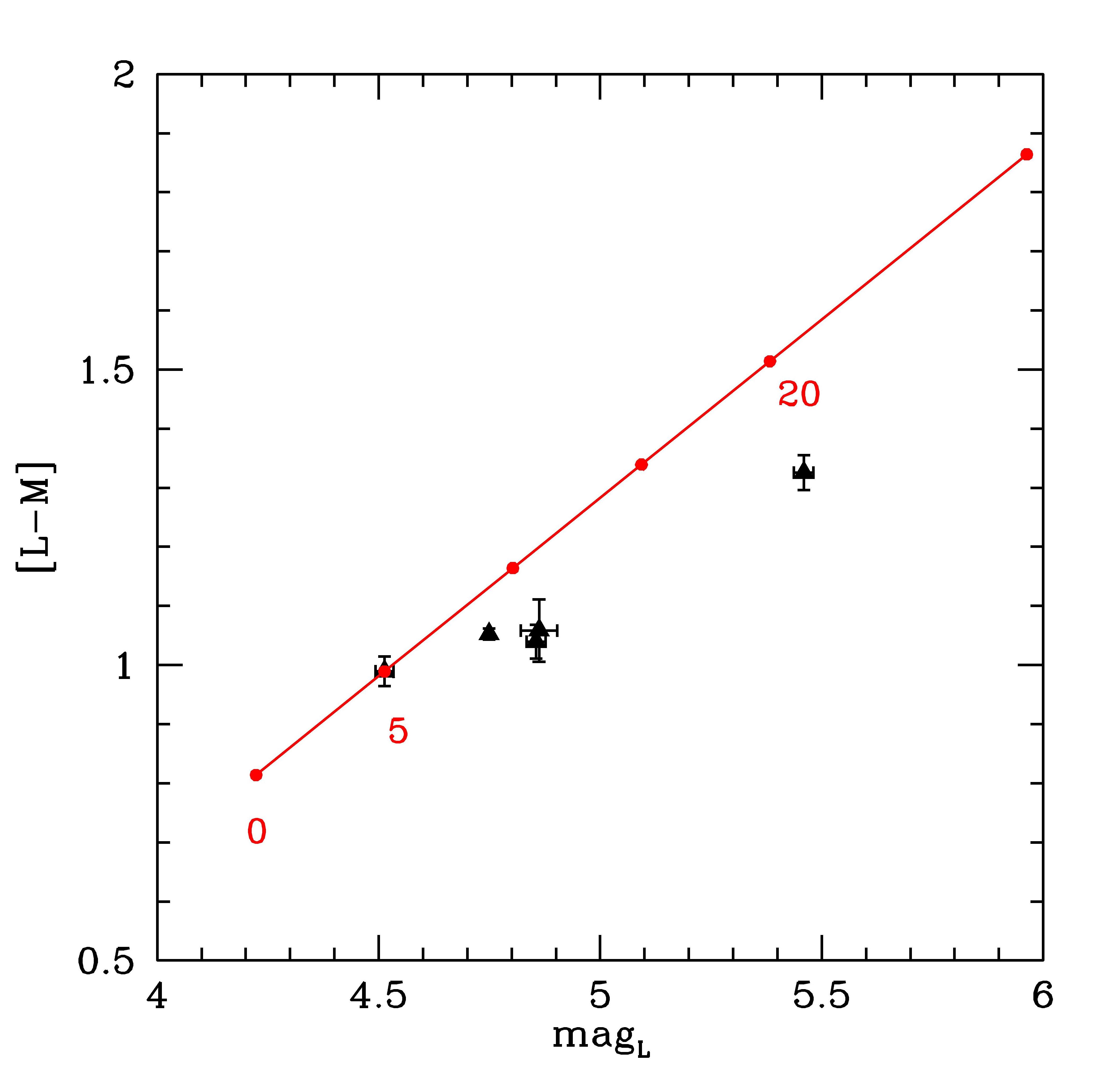}
\caption{\label{col_LM:fig} PV Cep color-magnitude plot ([L-M] versus mag L).
Solid line represents the reddening law by Rieke \& Lebofsky (1985), where 
different values of A$_V$ are indicated by solid dots.}
\end{figure}

Photometric data collected in the L and M bands (see Table~\ref{variab:tab}) are presented in Figure~\ref{col_LM:fig} in terms of a color-magnitude
plot. The extinction vector (Rieke \& Lebofsky 1985), depicted as a red line, has been arbitrarily superposed to the data in a way such as to overlap the value of A$_V$ = 5 mag with the photometric point corresponding to the brightest status of PV Cep. We note two facts: ({\it i}) comparing the mid-IR variations with the extinction vector, we conclude they cannot be extinction driven, thus confirming what we already inferred in the near-IR (Paper I); ({\it ii}) the source becomes bluer while brightening, a behaviour that is typical of the EXor eruptive variables (Lorenzetti et al. 2012 - Paper II).

\begin{figure}
\includegraphics[width=14cm]{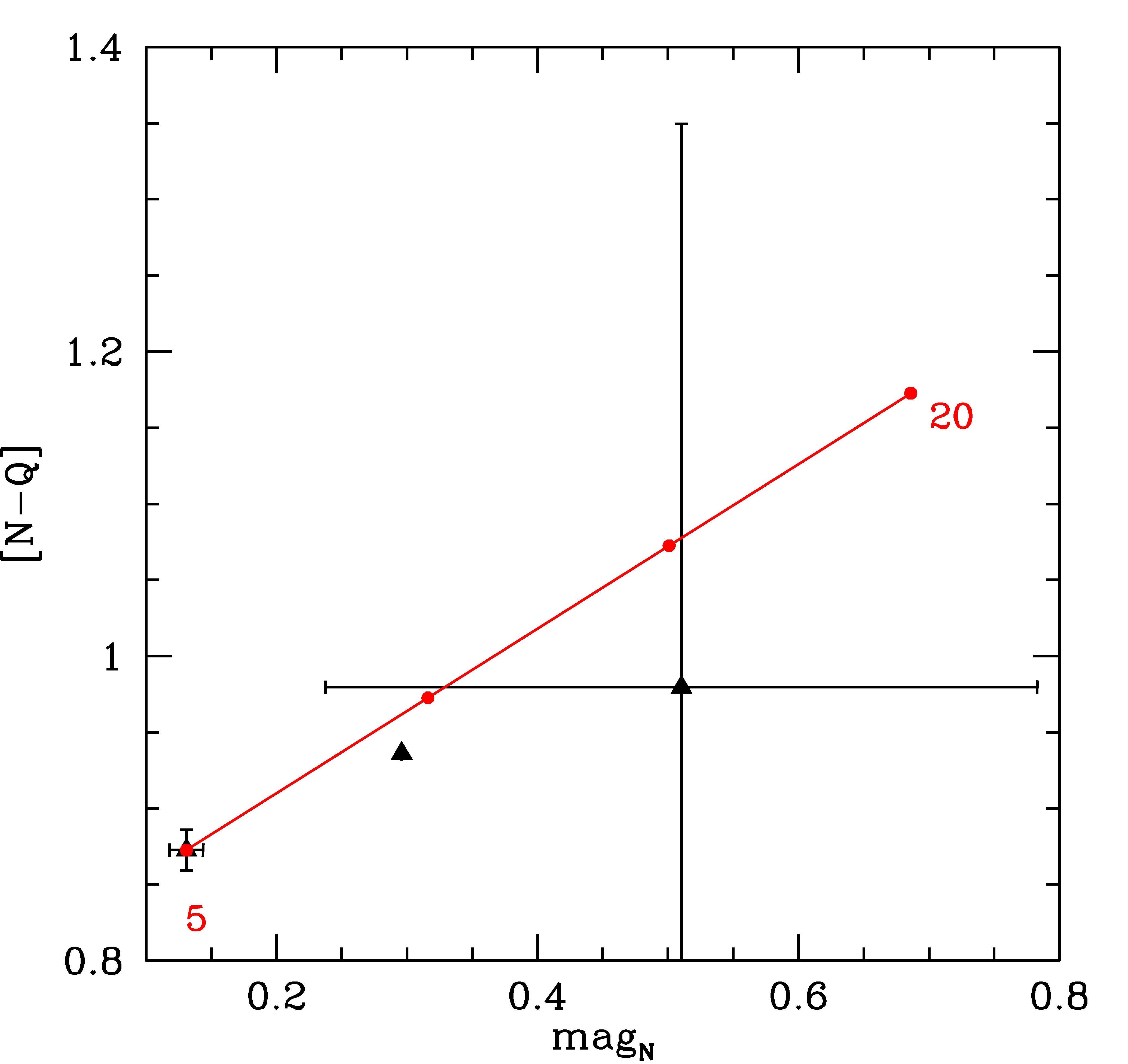}
\caption{\label{col_NQ:fig} PV Cep color-magnitude plot ([N-Q] versus mag N).
Solid line represents the reddening law by Rieke \& Lebofsky (1985), where 
different values of A$_V$ are indicated by solid dots.}
\end{figure}

As noted for the near-IR variations (Sect. 2), also the fluctuations depicted in Figure~\ref{col_LM:fig} could be interpreted not correctly if
completely attributed to a visual extinction variations between 5 and 20 mag. Although with a very poor statistics the same behaviour is recognizable in the [N-Q] versus N color-magnitude plot reported in Figure~\ref{col_NQ:fig}. 
A further case of interest is offered by mid-IR low resolution spectroscopy of the source. This was obtained twice by means of the facilities on board the satellites IRAS and ISO. ISO spectroscopy provided evidence for a 9.7 $\mu$m silicate absorption band and PAH emission at 11.2 $\mu$m (Acke \& van den Ancker 2004). In fact, a silicate absorption band had been previously detected in the IRAS-LRS spectrum (Chen, Wang \& He 2000) as well, although this feature was significantly deeper than the one observed by ISO. 
These absorption bands formally correspond to an optical depth $\tau_{9.7}$({\it IRAS}) = 1.8 and $\tau_{9.7}$({\it ISO}) = 0.3. According to the relationship A$_V$ (mag) = 16.6 $\times$ $\tau_{9.7}$ (Rieke \& Lebofsky 1985), the different measurements of $\tau_{9.7}$ correspond in turn to A$_V$ values of about 30 and 5 mag, respectively. Since line absorption might be related to the column material, which is independent on the beam-size for an illuminating point-like source, the detected difference in A$_V$ would seem to support the presence of local intrinsic extinction variations. However, such range of A$_V$ variability is not compatible with the fact that PV Cep remains visible at optical wavelengths even during its quiescence periods. 

In sum, all these considerations indicate that interpreting the PV Cep fluctuations in terms of an A$_V$ variation of about 20-25 mag is most likely not correct, so that an alternate mechanism is required, which would reproduce the observed variation of the optical and near-IR colors and change in the silicate absorption depth. In accordance with our conclusion, K11 remark how the wavelength dependence of the mid- and far-IR variations of PV Cep is different from that expected for dust extinction.

\subsubsection{SED modeling}

The color-magnitude plot depicted in Figure~\ref{col_LM:fig} indicates that the L-band fluctuations of PV Cep were sampled between a lower 
(corresponding to about 5.5 mag) and a higher (around 4.5 mag) state. Following the approach described in Paper II, we tried to investigate what happened between these two levels of brightness, checking if the difference between the SED relative to the brighter state and that of the fainter one can be fitted with a single black-body (BB). 
If this were the case, the appearance of a single temperature associated to the outburst event could effectively account for the observed fluctuation. To perform our check, we need two SEDs (outburst and quiescence) sampled at least in four photometric bands so as to make our two-parameter fit (see below) statistically meaningful. A conventional error of $\pm$10\% was arbitrarily assigned to any flux in order to take into account, in addition to the formal errors, also the effects due to differences in the adopted band-passes, effective wavelengths, and photometric systems. 
Among the cases presented for PV Cep
(see Table~\ref{variab:tab}), the two for which we have been able to reconstruct a SED sampled in four bands (H, K, L, and M) are that prior to 1984, considered as the high state, and that on Nov.~26, 2006, the low state. The H, K photometry was obtained simultaneously to L, M for the first epoch (H = 8.3 mag, K = 6.8 mag, Neckel \& Staude 1984), and two weeks later for the second epoch (H = 9.2 mag, K = 7.5 mag, our unpublished photometry).\

\begin{figure}
\includegraphics[width=14cm]{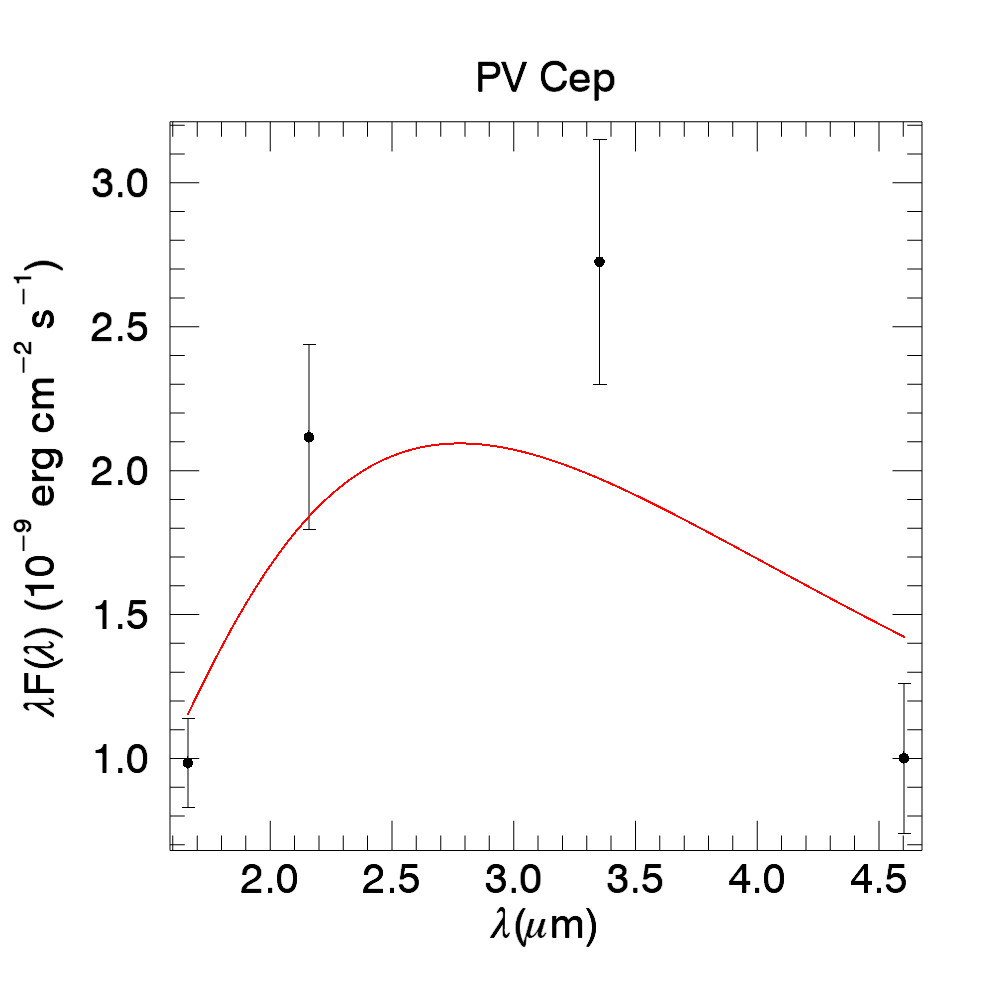}
\caption{\label{fit:fig} Best fit to the SED difference between high and low states of PV Cep. The curve represents a single temperature BB,
   whose value is given in Table~\ref{fit:tab}.}
\end{figure}

The free parameters of our simple model are $T$, the BB temperature, and $R$, the radius of the emitting region, while for
the extinction value A$_V$ we assumed 5 mag and 7 mag for the lower and higher state, respectively, which are the limits determined by all previous optical observations (see Sect.2). The fit of the differential SED is
depicted in Figure~\ref{fit:fig}, while the relevant parameters are given in Table~\ref{fit:tab}. As also noticed in Paper II, we point out that the shape of the true emitting area (that we do not model) is likely very different from the assumed uniform disk: for example it could
be the inner wall of the circumstellar disk or anything else: the radius $R$, derived by the fit, simply provides an indication of the length scale of the emission region.\

\begin{figure}
\includegraphics[width=16cm]{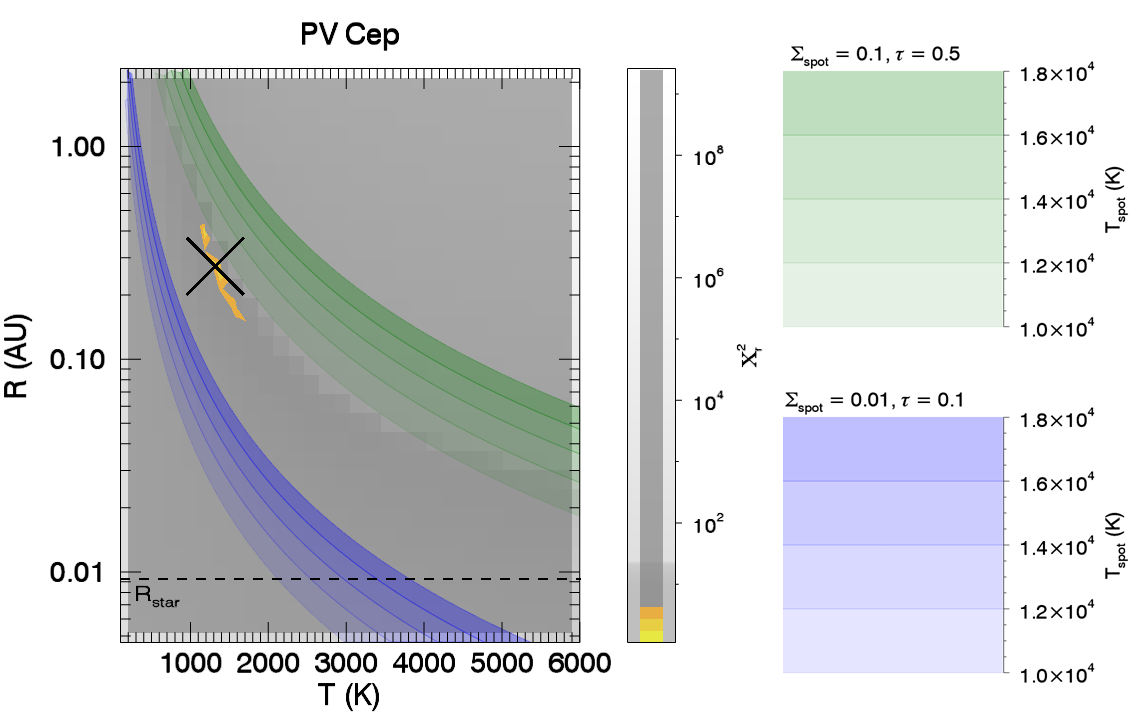}
\caption{\label{maps:fig} $\chi^2$ map (grey-scale) projected on the ($T$, $R$) parameters plane with superposed 1$\sigma$ to 3$\sigma$ confidence contours (in yellow and orange, respectively). The dashed horizontal line indicates the stellar radius (assumed 2 R$_{\odot}$), whereas the shaded stripes show the parameters region allowed by the spot model and the energy conservation in a range of spot temperatures T$_{spot}$ (10000 K to 18000 K) for two values (10$^{-3}$ and 5~10$^{-2}$ in blue and green, respectively) of the $\tau\cdot\Sigma_{spot}$ product (see text). The color bars relative to these two values, for different spot temperatures, are given in the two right panels. The black cross indicates the ($T$, $R$) pair corresponding to the absolute minimum of $\chi^2$.}
\end{figure}

Being the fit acceptable (reduced $\chi^2$ = 1.08), we should consider the hypothesis that an emitting region (assumed to be a uniform disk) at about 1300 K, with a radius R of about 0.27 AU, appeared as an additional component while the source moved from the low to the high state we have considered. The scale length of 0.27 AU is typical of the inner portions of the circumstellar disks of Herbig Ae stars (Vinkovi\'{c} 2006), as PV Cep is often classified.
We can explain the increase of $T$ in PV Cep in the same scenario we demonstrated to be valid to describe all the EXor outbursts (Paper II), and adopting for the spot emission the same model by Calvet \& Gullbring (1998) which predicts spot temperatures ($T_{spot}$) between 10,000 and 18,000 K. Since the appearance of this kind of black-body has been never observed, we assume that the energy irradiated by the spot equals that emitted by the surrounding region.
The power emitted by the spot linearly depends on two factors: ({\it i}) the time interval $\tau$ during which the spot temperature remains at its maximum value, expressed as a fraction of the observed bright state duration, and ({\it ii}) the spot surface $\Sigma_{spot}$, expressed as a fraction of the stellar surface. Given such linearity, we treated the product $\tau \cdot \Sigma_{spot}$ as a single variable. 
Figure~\ref{maps:fig} depicts the results of our analysis. Two extreme values of the product $\tau \cdot \Sigma_{spot}$ are considered:
0.001 ($\tau$ = 0.1 $\times$ $\Sigma_{spot}$ = 0.01) and 0.05 ($\tau$ = 0.5 $\times$ $\Sigma_{spot}$ = 0.1), indicated in blue and green, respectively. The considered upper value for $\tau$ is compatible with the typical light curve of EXors (see e.g. V1118 Ori - Audard et al 2010) where the maximum flux level lasts about half of the total bright state duration. The lower value for $\Sigma_{spot}$ is compatible with the sizes predicted by Calvet \& Gullbring (1998). In Figure~\ref{maps:fig}, the area of intersection between the
$\chi^2$ minima and model predictions represent the ranges of $T$ and $R$ for which our simple spot model can account for the observations.
The derived ranges are depicted as confidence contours (in yellow-orange) superposed on the $\chi^2$ maps and are listed in Table~\ref{fit:tab}. 
The event considered here is different from that presented in Paper II for the same source,
nevertheless it is still associated with the appearance of a single black-body (and not with a stratification of temperatures), so reinforcing 
conclusions given in Paper II. These both events, maybe fortuitously, are associated to black-bodies with very similar temperatures 
(T = 1320 K instead of 1550 K) and very compact size (R = 0.14 AU and 0.27 AU). 
Noticeably,  minimum $\chi^2$ corresponds to $\Sigma_{spot}$ = 0.1 and $\tau$ = 0.5, for a spot temperature of 15,000 K.
In practice, a very reasonable spot having an area of about 10\% of the stellar surface, persisting about 50\% of the bright state duration, and with a temperature well compatible with the PV Cep mass accretion rate, is able to account for the observed behaviour.\\ 

In this scenario, it is reasonable to expect that the energy input would induce a variable silicate emission feature from the outer disk, which could explain the observed variation in the silicate absorption band.

\subsection{Far-IR behaviour}

Concerning the far-IR detections at 60, 100, and 170 $\mu$m, no significant variability can be recognized above a 3$\sigma$ level. Analysing a fading
phase, also K11 found the range 10-60 $\mu$m unaffected by variability, although  they noted a concordant dimming in the 70-90 $\mu$m region.
The three observations (see bottom of Table~\ref{variab:tab}) that cover all the three band-passes allow us to estimate the [60-100] and [100-170] 
colors of PV Cep, where we define the color as 
[$\lambda_1 - \lambda_2$] = log($F_{\lambda_1}/F_{\lambda_2}$)).
The colors derived from the three observations are not compatible with the presence of a single temperature black-body, but rather correspond to a stratification of temperatures (see Fig.~2 of Elia et al. 2005).
Hamidouche (2010) was able for the first time to resolve the PV Cep disk by means of sub arc-second interferometry at $\lambda$ = 1.3 and 2.7 mm. He derived a quite high value for the exponent of the dust absorption coefficient in the relation $\kappa \sim \lambda^{-\beta}$, obtaining $\beta$ = 1.75 (which is a value indicative of small dust grains) and inferred the presence of a quite massive disk of 0.8 $M_{\sun}$. Far-IR
fluxes (given in Table~\ref{variab:tab}) are fully compatible (we suspect they are exactly the same he used) with the spectral energy distribution
modelled by Hamidouche (2010 - its Figure 8). To summarize, the far-IR measurements collected here are consistent with the presence of a young massive disk mainly composed by unprocessed (small grains of) dust.

\section{Final remarks}

The outbursting young object PV Cep has been observed in the mid- far-IR range ($\lambda$=3-170$\mu$m) several times. 
Mid-IR photometry ($\lambda$=3-25$\mu$m) is affected by a certain degree of variability whose modality confirms what already suggested by optical/near-IR monitoring observations, namely that variable accretion, more than variable extinction, is the dominant mechanism. 
Far-IR photometry ($\lambda$=60-170$\mu$m) does not seem affected
by a significant variability, and speaks in favour of a stratification of temperatures likely occurring in a very young massive circumstellar disk.

%

\newpage

\scriptsize
\begin{table}
\begin{center} 

\caption{[L band - 170 $\mu$m] photometry of PV Cep. \label{variab:tab}} 
\medskip
{
\begin{tabular}{l|cc|cc}
\hline
\medskip
Epoch           &        \multicolumn{2}{c}{L band}               &       \multicolumn{2}{c}{M band}          \\ 
                &  (Jy)           & [$\lambda_{eff}$] - Ref       &    (Jy)       &  [$\lambda_{eff}$] - Ref  \\ 
\hline                  
$<$ 1984$^a$    & 4.0             &  [3.8] - 1                    & 5.1             &  [4.6] - 1                \\
Jun 96 - Jan 98 & ...             &    ...                        & 4.4 $\pm$ 1.1   &  [4.8] - 2                \\
Oct 29, 2004    & 4.40 $\pm$ 0.09 &  [3.6] - 3                    & 7.0 $\pm$ 0.1   &  [4.5] - 3                \\ 
Nov 26, 2006    & 1.84 $\pm$ 0.04 &  [3.6] - 3                    & 3.99 $\pm$ 0.08 &  [4.5] - 3                \\ 
Sep 16, 2009    & 3.54 $\pm$ 0.03 &  [3.6] - 3                    & 5.97 $\pm$ 0.02 &  [4.5] - 3                \\
Sep 17-22 2009$^b$ & 3.21 $\pm$ 0.07 & [3.6] - 3                   & 5.35 $\pm$ 0.10 &  [4.5] - 3                \\    
Jan 09-15 2010$^b$ & 3.19 $\pm$ 0.13 & [3.6] - 3                   & 5.41 $\pm$ 0.18 &  [4.5] - 3                \\  
\hline
\end{tabular}}
\end{center}
\end{table} 
\addtocounter{table}{-1}
   

\begin{table}
\begin{center} 
\caption{Continued. \label{variab:tab}} 
\medskip
{
\begin{tabular}{l|cc|cc}
\hline
\medskip
Epoch           &        \multicolumn{2}{c}{N band}               &       \multicolumn{2}{c}{Q band}          \\ 
                &  (Jy)          & [$\lambda_{eff}$] - Ref        &    (Jy)       &  [$\lambda_{eff}$] - Ref  \\                   
\hline
1984            &13.27 $\pm$ 0.03&   [12] - 5                     &33.18 $\pm$ 0.04&  [25] - 5                \\ 
1996            & 13.4           &   [12] - 6                     & 23.4           &  [25] - 6                \\
Jun 96 - Jan 98 & 11 $\pm$ 3     &   [12] - 2                     & 20 $\pm$ 5     &  [25] - 2                \\
Feb 26, 2007    &  ...           &     ...                        & 27 $\pm$ 1     &  [24] - 3                \\  
2010/2011       & 15.6 $\pm$ 0.2 &   [12] - 4                     &25.7 $\pm$ 0.1 &   [22] - 4                \\     
\hline

\end{tabular}}
\end{center}
\end{table}   
\addtocounter{table}{-1} 

\scriptsize
\begin{table}
\begin{center} 

\caption{Continued. \label{variab:tab}} 

{
\begin{tabular}{l|cc|cc|cc}
\hline
\medskip
Epoch    &   \multicolumn{2}{c}{60 $\mu$m}    &   \multicolumn{2}{c}{100 $\mu$m}    &   \multicolumn{2}{c}{170 $\mu$m}    \\ 
         &  (Jy)    & [$\lambda_{eff}$] - Ref &  (Jy)    &  [$\lambda_{eff}$] - Ref &  (Jy)    &  [$\lambda_{eff}$] - Ref \\                   
\hline
1984            &51.87 $\pm$ 0.06&   [60] - 5 & 51.2 $\pm$ 0.2 &  [100] - 5         &   ...           &  ...              \\     
1996            & 40.3           &   [60] - 6 & 51.0           &  [100] - 6         & 56.1            &  [170] - 6        \\
Jun 96 - Jan 98 & 37 $\pm$ 9     &   [60] - 2 & 37 $\pm$ 9     &  [100] - 2         & 50 $\pm$ 13     &  [170] - 2        \\ 
May 06 - Aug 07 & 51 $\pm$ 2     &   [65] - 7 & 39 $\pm$ 2     &   [90] - 7         & 46 $\pm$ 2      &  [160] - 7        \\     
Feb 26, 2007    & 35 $\pm$ 2     &   [70] - 3 &   ...          &      ...           &   ...           &   ...             \\

\hline\hline

\end{tabular}}
\end{center}
\medskip

- References to the Table: (1) Neckel \& Staude 1984; (2) \'{A}brah\'{a}m et al. 2000; (3) Kun et al. 2011; (4) {\it WISE} Catalog; (5) Weaver \& Jones 1992; (6) Elia et al. 2005; (7) {\it AKARI} Catalog; \\

$^a$ - This date (prior to 1984) is not specified in the original paper and refers to an observation in J, H, K, L, and M bands conducted in the same date.\\
$^b$ - Photometric data taken during the indicated dates have been averaged: the mean value and the standard deviation are given.
 
\bigskip

\end{table}

\normalsize
\begin{table}
\begin{center} 

\caption{Adopted and derived parameters for PV Cep. \label{fit:tab}} 
\bigskip

{
\begin{tabular}{l|c}
\hline
\medskip
Parameter                  &      Value                 \\ 
\hline
Distance                   & 500 pc                     \\
A$_V$ (higher state)       & 7.0 mag                    \\
A$_V$ (lower state)        & 5.0 mag                    \\
fit reduced $\chi^2$       & 1.08                       \\
BB temperature $T$         & 1320$^{+32}_{-238}$ K      \\
Radius (emit.region) $R$   & 0.27$^{+0.05}_{-0.04}$ AU  \\
Spot surface $\Sigma$      & 10\% stellar surface       \\
Spot duration $\tau$       & 50\% outburst duration     \\                        
\hline\hline

\end{tabular}}
\end{center}
\medskip

\end{table}   


\end{document}